\documentstyle[11pt,psfig,newpasp,twoside]{article}
\index{central compact objects}
\index{XMM-Newton}

\def\pdot {\dot P}

\def\ltsima{$\; \buildrel < \over \sim \;$}
\def\lsim{\lower.5ex\hbox{\ltsima}}

\def\gsim{\lower.5ex\hbox{\gtsima}}

\def\ee {1E~1207.4--5209~}
\def\cha {\textit{Chandra~}}

\def\xmm  {\textit{XMM-Newton~}}

\def\edcomment#1{\iffalse\marginpar{\raggedright\sl#1\/}\else\relax\fi}
\reversemarginpar

\begin{document}
\title{Out of the Chorus Line : What Makes 1E1207.4-5209 a Unique Object?}
 \author{Andrea De Luca, Patrizia Caraveo, Sandro Mereghetti, Michele Moroni}
\affil{Istituto di Astrofisica Spaziale e Fisica Cosmica, Sezione
di Milano  ``G.Occhialini'' - CNR, Via Bassini 15, I-20133 Milano,
Italy}
\author{Giovanni Bignami }
\affil{Centre d'Etude Spatiale des Rayonnements, Toulouse, France
\\ Universit\`a degli Studi di Pavia,
Via Ugo Bassi, 6 - Pavia, Italy }
\author{Roberto Mignani }
\affil{European Southern Observatory, D-85740, Garching, Germany}

\begin{abstract}
The discovery of deep spectral features in the X-ray spectrum of
1E1207.4-5209 has pushed this Isolated Neutron Star (INS) out of
the chorus line, since no other INS has shown significant
features in  its X-ray continuum. On August 2002, \xmm
devoted a two-orbit TOO observation to this target with the aim
to better understand the nature of  such spectral features, using
much improved statistics. Indeed, the 260 ksec observation
yielded  360,000 photons from 1E1207.4-5209, allowing for a very
sensitive study of the temporal and spectral behaviour of this
object.
\end{abstract}

\section{Introduction}
Neutron star atmosphere models predicted the presence of
absorption features depending on  atmospheric composition, but high
quality spectra, collected both by \cha and by \xmm, did
not yield  evidence for any feature (see Pavlov et al. 2002a
and Becker and Aschenbach 2002 for recent reviews). 
INS spectra are well fitted
by one or more black-body curves with, possibly, a power law
contribution at higher energies, but with no absorption or
emission features. \\ The spectrum of 1E1207.4-5209, on the
contrary, is dominated by two broad absorption features seen, at
0.7 and 1.4 keV, both by \cha (Sanwal et al, 2002)  and
\xmm (Mereghetti et al, 2002). To better understand the
nature of such features, \xmm devoted two orbits, for a
total observing time of 257,303 sec, to 1E1207.4-5209. In the two
MOS EPIC cameras the source yielded  74,600 and 76,700 photons in
the energy range 0.2 - 3.5 keV, while the pn camera recorded
208,000 photons, time-tagged to allow for timing studies. Analysis
of this long observation, while confirming  the  two
phase-dependent absorption lines at 0.7 and 1.4 keV, unveiled a
statistically significant third line at $\sim$2.1 keV, as  well
as  a  possible fourth  feature at  2.8 keV.  The nearly 1:2:3:4
ratio  of the line centroids,  as well as  the phase variation,
naturally following the pulsar B-field rotation, strongly suggest
that such lines are due to cyclotron resonance scattering
(Bignami et al. 2003).  A recent software release, based on a
better characterization of the EPIC instrument, prompted us to
revisit the data. While the spectral analysis results confirm and
strengthen the conlusions of Bignami et al. (2003), the temporal
and spatial analysis yielded  interesting new results which we
shall briefly outline (see De Luca et al. 2003 for details).

\section{Timing Analysis}

After converting the arrival times of the 208,000 pn photons to
the Solar System Barycenter, we searched the period range from
424.12 to 424.14 ms using both a folding algorithm with 8 phase
bins and the Rayleigh  test. The best period value and its
uncertainty (P = 424.13076$\pm$0.00002 ms) were determined
following the procedure outlined in Mereghetti et al. (2002).
Comparing the new period measurement of \ee with that obtained
with \cha in January 2000 (Pavlov et al. 2002b), we obtain a
period derivative $\pdot$=(1.4$\pm$0.3)$\times$10$^{-14}$ s
s$^{-1}$. However, Fig.1 (left panel) shows that the $\pdot$ value rests
totally on the first  \cha period measurement. Using only the
3 most recent values, the period derivative is unconstrained.
Thus, we cannot exclude that the observed spin-down, based on
only a few sparse measurements, be affected by glitches or
Doppler shifts induced by orbital motion. Questioning the
object's $\pdot$ would have far reaching consequences for the
understanding of 1E1207.4-5209 since the serious discrepancy
between the pulsar characteritic age ($\tau_{c}\sim5\times10^5$ yrs)
and the SNR age ($\tau_{SNR}\sim7$ kyrs) is entirely
based on the value inferred from the measurements summarized in
Fig.~1 (left panel).

\begin{figure}
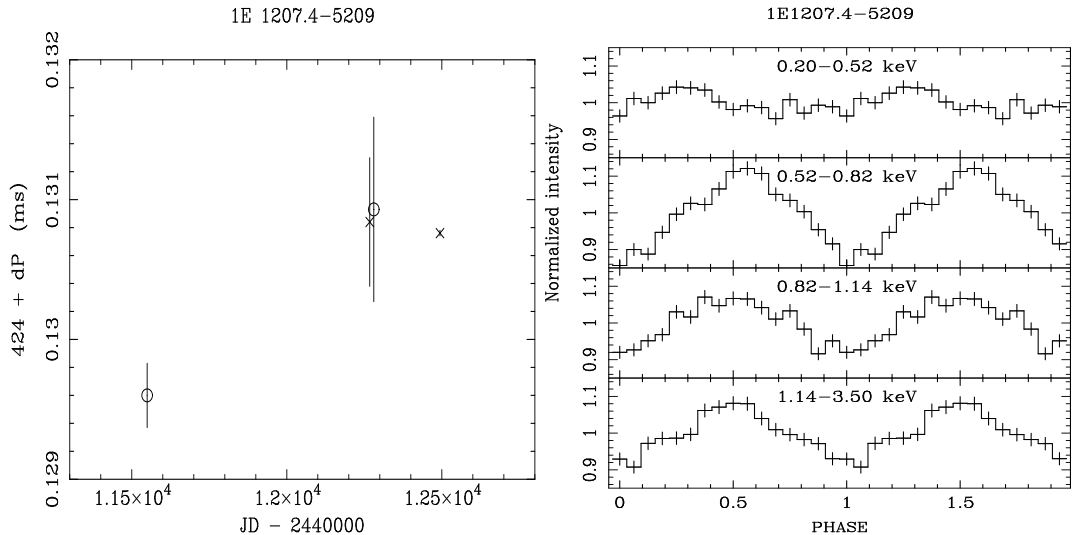

\centerline{\psfig{file=bignamig1_1.ps,width=7cm,height=7cm,angle=-90}
\psfig{file=bignamig1_2.ps,width=7cm,height=7cm,angle=-90}}
\caption{(Left) Period history of \ee . Circles are the \cha
measurements, crosses \xmm. (Right) Folded light curve of \ee in four energy ranges. Fitting
the pulse profiles with a function consisting of a constant plus a
sinusoid, we obtain pulsed fraction values of (2.7$\pm$0.7)\% in
the 0.2-0.52 keV energy range, (11.1$\pm$0.7)\% for 0.52-0.82 keV
, (7.1$\pm$0.7)\% for 0.82-1.14 keV, and (7.1$\pm$0.7)\% 1.14-3.5
keV. }
\end{figure}

To study the energy dependence of the pulse profile, we divided
the data in four channels with approximately 52,000 counts each:
0.2-0.52 keV, 0.52-0.82 keV, 0.82-1.14 keV and 1.14-3.5 keV. The
pulse profiles in the different energy ranges (Fig.~1, right panel)   show a
broad, nearly sinusoidal shape, with a pulsed fraction varying
from $\sim$ 3 to $\sim$ 11 \%  in the four energy intervals.  It
is worth noting that the minimum pulsed fraction is found in the
0.20-0.52 keV energy range, the only portion of the spectrum free
from absorption lines. Indeed, Fig.~1 (right panel) is an independent
confirmation of the findings of Bignami et al (2003) who ascribed
the source pulsation to the absorption lines phase variation.

Finally, comparing the shapes of the light curves of Fig.~1 (right), we
see for the first time a phase shift of nearly 90$^{\circ}$
between the profile in the lowest energy range ($<$0.52 keV) and
those at higher energies.


\section{Optimizing the X-ray position}

To derive the sky coordinates of \ee we computed independently
for the MOS1 and MOS2 cameras the boresight correction to be
applied to the default EPIC astrometry. We used the Guide Star
Catalog II
(GSC-II\footnote{http://www-gsss.stsci.edu/gsc/gsc2/GSC2home.htm})
to select, amongst our $\sim$200 serendipitous detections, 6
sources with a stellar counterpart to be used to correct the EPIC
astrometry. The rms error between the refined X-ray and GSC-II
positions is $\sim$1 arcsec per coordinate. The resulting MOS1
position of \ee is $\alpha_{J2000}=12h10m00.91s$,
$\delta_{J2000}=-52^{\circ}26'28.8"$ with an overall error radius
of 1.5 arcsec. The MOS2 position is $\alpha_{J2000}=12h10m00.84s$,
$\delta_{J2000}=-52^{\circ}26'27.6"$, with an uncertainty of 1.5
arcsec, fully consistent with the MOS1 coordinates.

\section{Search for the optical counterpart}

The field of 1E 1207$-$5209 was observed with the 8.2-meter UT-1
Telescope (Antu) of the ESO VLT (Paranal Observatory).
Observations were  performed with the the FOcal Reducer and
Spectrograph 1 (FORS1) instrument. Images   were   acquired
through   the   Bessel   $V$ and $R$  filters  for  a  total
integration  time of  $\sim $2  and 3 hrs,  respectively.  Fig.~2
 shows the inner portion  of the combined FORS1
$V$-band image centered on the target position, with the MOS1 and
MOS2 error cicles superimposed.    A faint object (marked with
the  two ticks in Fig.~2) is detected just outside the southern
edge of the MOS1 error circle and showed variability along the
time span covered by our observations. In any case, its position falls  more
than 2 arcsec  away from the intersection  of the MOS1/MOS2 error
circles, which we  regard as the most probable region. \\No
candidate counterpart  is detected within the  MOS1/MOS2 error
circles down to R$\sim$27.1 and V$\sim$27.3, which we assume as
upper limits on the optical flux of 1E 1207$-$5209. 
For the  X-ray derived interstellar  absorption ($A_{V} =0.65$)  and a
distance of 2 kpc, the measured u.l. rule out any hypothetical ``normal'' stellar 
companion other than a very low-mass main sequence star.  If we assume
that 1E  1207-5209 is indeed {\it  isolated}, we can  derive a neutron
star optical luminosity $\le 3.4 ~ 10^{28}$ erg s$^{-1}$ or $\le 4.6 ~
10^{-6}$ of  its rotational energy loss,  a value similar  to those of
middle-aged INSs.  Since the VLT flux upper limits are $\ge
100$ higher  than the extrapolation of  the \xmm~ blackbody  (De Luca et
al.  2003), they can not constrain the optical spectrum.  


\begin{figure}
\centerline{\psfig{file=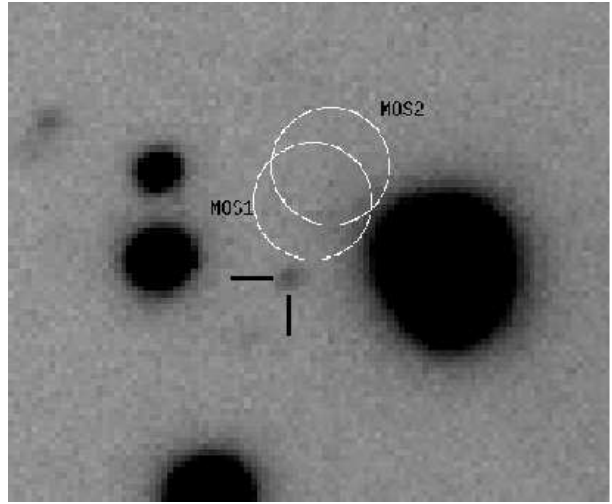,width=8cm,angle=-90}}
\caption{FORS1 $V$-band image centered on the  target position,
with the MOS1  and MOS2 error circles (2 arcsec radius) superimposed.}
\end{figure}

\end{document}